\documentstyle{article}

\newcommand{\be}{\begin{equation}		}
\newcommand{\ee}{\end{equation}  }
\newcommand{\bea}{\begin{eqnarray}		}
\newcommand{\eea}{\end{eqnarray}  }

\font\tenbf=cmbx10
\font\tenrm=cmr10
\font\tenit=cmti10
\font\elevenbf=cmbx10 scaled\magstep 1
\font\elevenrm=cmr10 scaled\magstep 1
\font\elevenit=cmti10 scaled\magstep 1

\renewenvironment{thebibliography}[1]
 { \elevenrm
   \begin{list}{\arabic{enumi}.}
    {\usecounter{enumi} \setlength{\parsep}{0pt}
     \setlength{\itemsep}{3pt} \settowidth{\labelwidth}{#1.}
     \sloppy
    }}{\end{list}}

\begin{document}

\begin{center}{{\tenbf PAST, PRESENT, AND POSSIBLE FUTURE LIMITS \\
               \vglue 3pt
               ON THE PHOTON REST MASS\\}

%\vglue 5pt
%{\ninerm (For 20\% Reduction to 8.5 $\times$ 6 in Trim Size)\\}

\vglue 1.0cm
{\tenrm MICHAEL MARTIN NIETO \\}
\baselineskip=13pt
{\tenit Theoretical Division, Los Alamos National Laboratory\\}
\baselineskip=12pt
{\tenit University of California, Los Alamos, NM 87545, U.S.A.\\}

\vglue 0.8cm
{\tenrm ABSTRACT}}
\end{center}
\vglue 0.3cm
{\rightskip=3pc
 \leftskip=3pc
 \tenrm\baselineskip=12pt
 \noindent
In an historical context, present limits on the photon rest
mass are reviewed.  More
stringent, yet speculative, limits which have been proposed are mentioned.
Finally,  new theoretical ideas and
 possible experimental improvements on the present limits are discussed, along
with possible relationships between these two areas.

 \vglue 0.6cm}
{\elevenbf\noindent 1. Historical Background}
\vglue 0.4cm
\baselineskip=14pt
\elevenrm

As recounted in reviews on the topic,$^{1,2}$  studies of the force law between
electric charges date back to the middle of the 1700's.  The first
experimental test of the inverse-square law was done in 1769 by
Robison, who was partially motivated by the work of Benjamin Franklin.
Although Robison's work, and also that of Cavendish, preceded that of Coulomb,
their results were not published until decades later.  Coulomb
published his results in a timely manner and hence received recognition.
The results of these authors were parametrized  in terms of a force law of the
form
\be
F \propto \frac{1}{r^{2+q}},
\ee
as was much later work.

However, with the discovery  of the quantum-mechanical generalization   of
Maxwell's equations,  we now know that the inverse-square law can be viewed as
a
consequence of the masslessness of the gauge particle which mediates the
electromagnetic force, the photon.  If the photon were to have a mass, then the
``massive Maxwell  equations," or Proca equations, are
\bea
{\bf \nabla \cdot  E} &=& 4\pi\rho - \mu^2 \Phi ,  \\
{\bf \nabla \times E} &=& -\frac{1}{c} \frac{\partial {\bf H}}{\partial t} ,\\
{\bf \nabla \cdot  H} &=& 0, \\
{\bf \nabla \times H} &=& \frac{1}{c} \frac{\partial {\bf E}}{\partial t}
      +\frac{4\pi}{c} {\bf J} - \mu^2 {\bf A} .
\eea
One still has the definitions
\bea
{\bf H} &=& {\bf \nabla \times A} , \\
{\bf E} &=& -{\bf \nabla}V
     -\frac{1}{c} \frac{\partial {\bf A}}{\partial t} .
\eea
The charge conservation condition is
\be
(\Box + {\mu}^{2})A_{\nu} = \frac{4\pi}{c}J_{\nu}.
\ee
One loses gauge invariance and instead must satisfy the Lorentz gauge
condition,
\be
\partial^{\lambda}A_{\lambda} = 0.
\ee
In the above, $\mu$ is the photon rest mass, in units of inverse length
$(c/\hbar)$.

\vglue 0.5cm
{\elevenbf\noindent 2.  Consequences of ${\bf \mu \neq 0}$}
\vglue 0.2cm
{\elevenit\noindent 2.1. The Velocity of Light}
\vglue 0.3cm

What would be the consequences of there being a photon mass?  The first is
that, as with all other massive particles, the velocity of light would not be
a constant, and $c$ would only be the ``limiting velocity."
(de Broglie pioneered this idea.)

In particular, the group velocity of light would be
\be
\frac{v_g}{c} = \left[1 - \frac{\omega_{\mu}^2}{\omega^2}\right]^{1/2} ,
\hspace{0.5in}  \omega_{\mu} = \mu c .   \label{omega}
\ee
The best limit on the photon mass using  this method comes from measuring the
difference in arrival times of different-frequency radio waves from the Crab
pulsar.$^3$  A dispersion is seen.   This dispersion could be taken to
indicate a photon mass if it were not for the fact that the electron plasma
in interstellar space produces the same type of dispersion relation as that of
Eq.
(\ref{omega}), but with $\omega_{\mu}$ replaced by
\be
  \omega_{p} = \left(\frac{4\pi n e^2}{m}\right)^{1/2},
\ee
where $n$ is the free electron density.  Because of other, better photon
mass limits, the result is interpreted as being due to an average, interstellar
density of $n=0.028$ electrons/cm$^3$.  But  the dispersion so obtained
implies,
by itself, a photon mass limit of
\be
\mu \leq 6 \times 10^{-12}\,eV = 10^{-44}\,gm = 3 \times 10^7\,cm^{-1}.
\ee

\vglue 0.2cm
{\elevenit\noindent 2.2. Coulomb's Law}
\vglue 0.3cm

If the photon has a mass, then the scalar potential is defined by
\be
(\nabla^2 - \mu^2)\Phi(r) = -4\pi \rho(r),
\ee
meaning it has a Yukawa form:
\be
\Phi(r) = e\frac{\exp[-\mu r]}{r} .
\ee

Now consider two concentric, conducting, spherical shells of radii $a$ and $b$,
that are first grounded, then decoupled from ground, and then have a potential
applied to the outer shell of radius $a$.  Then the above paragraph means that
the
electrostatic potential difference between the two shells is no longer zero.
Rather, it is
\be
\frac{\Delta V}{V} = \frac{\phi(a) - \phi(b)}{\phi(a)} \simeq
\frac{1}{6} \mu^2(a^2-b^2) + {\cal O}[(\mu a)^4] , \label{coulomb}
\ee
where
\be
\phi(r) = K\left[\frac{e^{\mu r} - e^{-\mu r}}{2\mu r}\right] .
\ee
In a sophisticated multi-shell version of this basic idea, Williams, Faller,
and Hill obtained the best laboratory limit to date on the photon mass,$^4$
\be
\mu \leq  10^{-14}\,eV = 2 \times 10^{-47}\,gm = 6 \times 10^{-10}\,cm^{-1}.
\ee

Before continuing, note the result of both Eqs. (\ref{omega}) and
(\ref{coulomb}), that physical effects of a photon mass first appear in order
$(\mu L)^2$, where $L$ is some scale size of the system.  This is a
theorem.$^1$  In particular, it means that to get a good limit, you either have
to make a very precise measurement with a small apparatus (as was done for
Coulomb's Law), or else, if the measurement is not as precise, one needs a very
large apparatus (as was done with the Crab nebula and with the magnetic
measurements of the next subsection).

\vglue 0.2cm
{\elevenit\noindent 2.3. Planetary Magnetic Fields}
\vglue 0.3cm

With a massive photon,  magnetic multipole fields will change to a Yukawa
form just as electric multipole fields do.  In particular, a field from a
magnetic dipole, $D$,  becomes
\be
{\bf H} = \left[\frac{D e^{-\mu r}}{r^3}\right]
\left[\left( 1 + \mu r+\frac{1}{3}\mu^2r^2\right)
     (3{\bf \hat{z}\cdot\hat{r}\hat{z}}-{\bf \hat{z}})
   -\frac{2}{3}\mu^2r^2{\bf \hat{z}}\right].
\label{jup}
\ee
In addition to the general ``Yukawa" contraction of the size of the field,
there is a new last term in Eq. (\ref{jup}), the ``external field
effect:"  $ -\frac{2}{3}\mu^2r^2\hat{z}$.  On a sphere surrounding the dipole,
it appears to be a constant field antiparallel to the dipole.  The first
person to look for such an effect was Schr\"{o}dinger, who studied the Earth's
magnetic field. (He was interested in a finite photon mass in
conjunction with his ideas to unify gravity and electromagnetism.)

The best application of this method to date, using higher multipoles in the
analysis, was done by Davis, Jr., {\it et al.}.$^5$  They considered an even
bigger magnet than the Earth,  Jupiter.  Using data from the Pioneer 10 flyby
of Jupiter, a limit of
\be
\mu \leq 6 \times 10^{-16}\,eV = 8 \times 10^{-49}\,gm =
2 \times 10^{-11}\,cm^{-1}
\ee
was obtained.

\vglue 0.5cm
{\elevenbf\noindent 3. More Speculative Limits}
\vglue 0.4cm

For many years now, more speculative limits on the photon mass have been
proposed based on considerations of distant astronomical objects with
magnetic fields spread over large volumes.   They range from $\sim
2\times 10^{-20}\,eV$, from the properties of the galactic magnetic field, to
$\sim 10^{-27}\,eV$, from the properties of interstellar gas in the Small
Magellanic Cloud.  (Consult Ref. 6 for a discussion of these limits.)

However, the magnetohydrodynamics of these distant, large objects is in no way
rigorously understood.  Such simple questions as if the plasma is driving the
field, or {\it visa versa}, remain subjects of debate.  It is not understood
why
these huge magnetized bodies can maintain their coherence over time scales
approaching the cosmological.  Indeed, this observation leads to the next
section.

\vglue 0.5cm
{\elevenbf\noindent 4. New Theoretical Ideas}
\vglue 0.2cm
{\elevenit\noindent 4.1. Strings}
\vglue 0.3cm

In normal point field theory, couplings such as $S A^{\lambda} A_{\lambda}$,
where $S$ is a scalar, are zero by gauge invariance.  However, Kosteleck\'{y},
Potting, and Samuel have pointed out that this is no longer true in string
theories.$^{7}$  In particular, if $S$ is the gravitational curvature, $R$,
then this coupling could lead to primeval magnetic fields of the sizes
presently observed.

\vglue 0.2cm
{\elevenit\noindent 4.2. Spontaneous Symmetry Breaking}
\vglue 0.3cm

With the successes of the standard model, it has been natural to ask if the
$U(1)$ of electromagnetism might be  broken at some low temperature,
thereby yielding a photon mass in that regime.  Although dynamical symmetry
breaking is ruled out,$^{8}$ much interest has been shown in possible
spontaneous
symmetry breaking.$^{9-12}$  This is true even though it is hard to find such a
theory which would predict a new experimental signature but yet would not
already
be in conflict with other experiment.   Further, there is no hard prediction of
what the transition temperature might be. Nonetheless, the idea is fascinating,
and has partially stimulated thoughts about one possible new experiment.

\vglue 0.5cm
{\elevenbf\noindent 5. Possible New Experiments}
\vglue 0.2cm
{\elevenit\noindent 5.1. Coulomb's Law}
\vglue 0.3cm

The experiment of Williams, Faller, and Hill$^4$ remains the best laboratory
limit on the photon rest mass.  Note that, because of the theorem we
mentioned,  any increased experimental sensitivity yields a better photon mass
limit as $(\mu L)^2$.  Therefore, one needs to improve the signal to noise
ratio
of an experiment by a factor of $100$ to get an improved photon mass limit of a
factor of $10$.  Thus, it will be a nontrivial task to go beyond the present
laboratory limit.

Even so, Henry Hill$^{13}$ is considering such an experiment, and has discussed
it with his two earlier collaborators.$^4$  To begin, a significant
improvement over the previous experiment appears possible just by advances
of standard experimental techniques.  Further,
another significant reduction in noise should be possible by  using  dilution
refrigeration technology to reduce the   temperature of the
apparatus to $mK$.  These cryogenic techniques  have been developed for
Weber-bar gravitational wave detectors.

This last would allow a first search, however theoretically ill-defined, for a
low-temperature phase transition.  One would have to try to reduce all
electrical signals to as low a level as possible, since signals could ruin a
phase transition, just as a magnetic field can ruin superconductivity via the
Meissner effect.$^{12}$

\vglue 0.2cm
{\elevenit\noindent 5.2. Solar System Magnetic Fields}
\vglue 0.3cm

Although there have been other missions to Jupiter since  Pioneer 10,
 for our purposes no striking improvement in data  has been obtained
since these missions have also been flybys.  However, that situation will
change with the arrival of the Galileo probe to Jupiter in 1995.$^{14}$  This
craft will have enough fuel to maintain attitude control  for approximately 10
eccentric orbits about Jupiter, with distances from the surface ranging from a
single closest perijove of $4\,R_J$ to a varying distance of about $100\,R_J$.
In
principle this added data could allow a a better Jupiter photon mass limit by a
factor of perhaps $2$ to $4$.  However, the closest approach of
$4\,R_J$ is further out than the Pioneer 10 distance of $2.84\, R_J$.

Even more intriguing is the Ulysses mission to the sun.$^{15,16}$  The probe
has
just encountered Jupiter,$^{16}$ obtaining a gravity boost that has set it in a
solar polar orbit.  During June-November 1994 and June-September 1995 Ulysses
will pass over the south and north polar regions at distances between $1.7$ to
$2.9$  $AU$.  A prime advantage of this orbit is that one expects the solar
wind
to be greatly reduced over the poles.  On the other hand, the nature of the
complications from the ``Archimedes spiral," caused by the  magnetic
field rotating with the sun, remain a subject of discussion.$^{17}$

Most significantly, the true time-dependent value of
the solar dipole magnetic moment is uncertain, but it is believed to be
approximately$^{18}$ $6-12$ $Gauss$ $R_S^3$.  Since an Astronomical Unit is
about
$200 R_S$, this means that one would expect the solar dipole field in these
regions to be approximately $0.01 \gamma$ ($1\,Tesla = 10^4\,Gauss = 10^9\,
\gamma$). This is just at the limit of what onboard magnetometers can measure.

Thus, although the
the sun is clearly the biggest magnet we can ``get our hands on," we will have
to
wait  to find out if Ulysses' orbit will
be close enough to the sun to obtain an improved solar-system limit on the
mass of the photon.

\vglue 0.5cm
{\elevenbf\noindent 6. Acknowledgements \hfil}
\vglue 0.4cm

First, I want to acknowledge Fred Goldhaber, with whom I worked in this field
for
many years.  I have always been happy to take credit for the efforts
of my friends.  Second, I  want to take this chance to thank
someone whom I never have in print:  T. Alexander Pond.  It was he who, back in
my Stony Brook days, first explained  Schr\"{o}dinger's work on this problem to
Fred and myself, which is what started the two of us off.  Finally, since it
was
Alan Kosteleck\'{y} who roped me into giving a talk at David Cline's Workshop,
I
guess I should also thank (blame?) him.  We hope to pursue this topic.

\newpage
%\vglue 0.5cm
{\elevenbf\noindent 6. References \hfil}
\vglue 0.4cm

\end{document}